\documentclass[12pt]{article}\usepackage[hyperfootnotes=false]{hyperref}
   \usepackage{epsfig}
      \usepackage{amsmath}
      \usepackage{amssymb}
\usepackage{cite}
  \usepackage{graphicx}
  \usepackage{color}
  \newlength{\abstractwidth}
  \renewcommand{\thefootnote}{\fnsymbol{footnote}}
  \renewcommand{\thanks}[1]{\footnote{#1}} 
  \newcommand{\starttext}{
  \setcounter{footnote}{0}
  \renewcommand{\thefootnote}{\arabic{footnote}}}
  \renewcommand{\theequation}{\thesection.\arabic{equation}}
  \newcommand{\be}{\begin{equation}}
  \newcommand{\bea}{\begin{eqnarray}}
  \newcommand{\eea}{\end{eqnarray}}
  \newcommand{\beq}{\begin{equation}}
  \newcommand{\ee}{\end{equation}}
  \newcommand{\eeq}{\end{equation}}

  \def\ba{\begin{eqnarray}}
  \def\ea{\end{eqnarray}}

  \def\12{{1 \over 2}}

 \def\simleq{\; \raise0.3ex\hbox{$<$\kern-0.75em
      \raise-1.1ex\hbox{$\sim$}}\; }
 \def\simgeq{\; \raise0.3ex\hbox{$>$\kern-0.75em
      \raise-1.1ex\hbox{$\sim$}}\; }

\def\O2{\Omega_2}

\def\bi{\begin{itemize}}
  \def\ei{\end{itemize}}

\def\sc{\setcounter{equation}{0}}

\def\W{$\Omega$}
\def\W'{$\Omega$}

\def\V{\Omega}
\def\V'{\Omega}

\def\nref#1{(\ref{#1})}

\def\la{\label}

\begin{document}
  \renewcommand{\theequation}{\thesection.\arabic{equation}}

\begin{titlepage}
  \rightline{}
  \bigskip

  \bigskip\bigskip\bigskip\bigskip

    \bigskip
\centerline{\Large \bf {  Quantum corrections to   }}
\bigskip
\centerline{\Large \bf {holographic entanglement entropy
}}
    \bigskip
\bigskip

  \begin{center}
 \bf {Thomas Faulkner$^1$, Aitor Lewkowycz$^2$ and Juan Maldacena$^1$}
  \bigskip \rm
\bigskip

$^1$ Institute for Advanced Study,  Princeton, NJ 08540, USA
\\ ~~
\\
$^2$ Jadwin Hall,  Princeton University, Princeton, NJ 08544, USA \\
\rm

\bigskip
\bigskip

\vspace{1cm}
  \end{center}

  \bigskip\bigskip

 \bigskip\bigskip
  \begin{abstract}

We consider entanglement entropy in quantum field theories with a gravity dual.
In the gravity description, the leading order contribution comes from the area
of a minimal surface, as proposed by Ryu-Takayanagi.
Here we describe  the one loop correction to this formula.
The minimal surface divides the bulk into two regions. The bulk loop correction is
essentially given by the bulk entanglement entropy between these two bulk regions.
We perform some simple checks of this proposal.

 \medskip
  \noindent
  \end{abstract}

  \end{titlepage}

    \starttext \baselineskip=17.63pt \setcounter{footnote}{0}
  \tableofcontents

  \sc

\section{ Holographic entanglement entropy  }

\begin{figure}[h!]
\begin{center}
\vspace{5mm}
\includegraphics[scale=1]{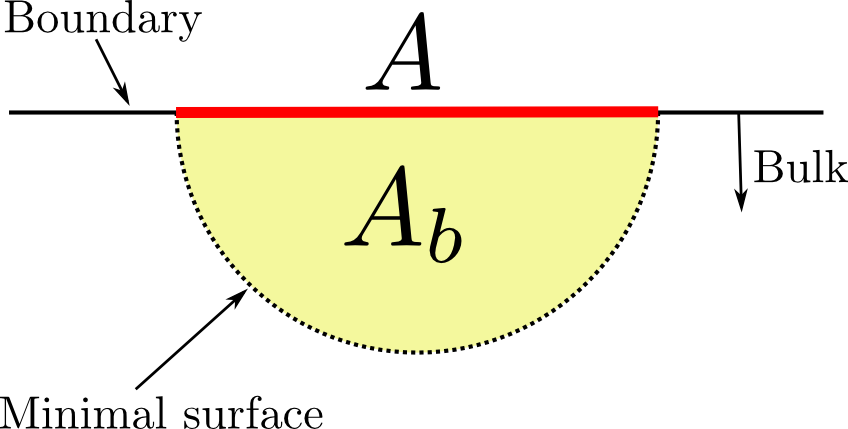}
\vspace{5mm}
\caption{
The red segment indicates a spatial region, $A$,  of the boundary
theory. The leading contribution to the entanglement entropy is computed by the area of a minimal surface that ends at the boundary of
region $A$. This surface divides the bulk into two, region $A_b$ and its complement. Region
$A_b$ lives in the bulk and has one more dimension than region $A$. The leading correction
to the boundary entanglement entropy
is given by the bulk entanglement entropy between region $A_b$ and the rest of the bulk.
  }
\label{RTBasic}
\end{center}
\end{figure}

In quantum field theories, it is interesting to
 compute the entanglement entropy among various
subregions. For example, we can consider a region $A$ and compute the entanglement entropy
between region $A$ and the rest of the system, see figure \ref{RTBasic}.
In theories with a gravity dual there is a very simple prescription for computing this
entropy \cite{RT,RT2}. We first find a minimal area surface that ends on the boundary of region $A$, at the
boundary of the bulk, see figure \ref{RTBasic}. Then the entropy is given by the area of this surface,
\be
\la{TRent}
S_{cl}(A) = { ({\rm Area} )_{\rm min} \over 4 G_N }
\ee
In situations where we can apply the replica trick, this formula was proven  for $AdS_3$ in \cite{Faulkner:2013yia,Hartman:2013mia} and more generally in \cite{gge}.
This is the correct result to leading order in the $G_N$ expansion.
If the boundary theory is a large $N$ gauge theory, then  \nref{TRent} is
 of order $N^2$. The leading term
\nref{TRent} comes from classical physics in the bulk. Here we consider the quantum corrections
to this formula. Namely, corrections that come from quantum mechanical effects in the bulk.
These are of order $G_N^0$  (or $N^0$).

We find  that the quantum corrections are essentially given by the bulk entanglement
entropy. More precisely, the minimal surface that appears in \nref{TRent} divides the bulk into
two regions. We denote by $A_b$ the bulk region that is connected to the boundary region $A$,
see figure \ref{RTBasic} .
Then the bulk quantum correction is essentially given by the bulk entanglement entropy  between
region $A_b$ and the rest of the bulk. Namely, at this order, we can think of the bulk as an effective
 field theory
living on a fixed background geometry and compute the entanglement entropy of region $A_b$ as
we would normally do in any quantum field theory\footnote{Caution: do not confuse the bulk entanglement entropy  \nref{bulk-ent} with the one computed by the area formula \nref{TRent}.
 Both are computed in the bulk and are entanglement entropies, so unfortunately we have a clash
of terminology. Hopefully, this will not cause confusion. Note also that
 \cite{Bianchi:2012ev} discussed a  proposal of entanglement entropy in gravitational theories which does not require the surfaces to be minimal.  } .  This is a computation in the bulk effective
field theory, it depends on the details of the bulk fields.
We can then write the quantum correction as
\begin{align}
   S(A) &= S_{cl}(A) + S_q(A) + \mathcal{O}(G_N)  ~,~~~~~~~~~ \label{sumaent}\\
  S_q(A) & =  S_{\rm bulk-ent}(A_b) + \cdots \label{bulk-ent}
\end{align}

The dots in \nref{bulk-ent}
 denote some extra one loop terms that can be expressed (like the classical term \nref{TRent})
as an integral of local quantities. We will give a more detailed discussion of these terms below. They
include terms that cancel the UV divergencies of the bulk entanglement entropy, so that
$S_q$ is a finite quantity. In the case of black holes,
this expression  for the quantum correction has been discussed
in
\cite{Bombelli:1986rw,Srednicki:1993im,Callan:1994py,Susskind:1994sm,Solodukhin:1994yz,Fursaev:1994ea,Solodukhin.BH,offvson},
 with increasing
degrees of precision.

We first present a sketch of an argument for this formula. We then consider various simple
checks.

\section{An argument }
\label{Argument}

In static situations one can use the replica trick to compute the entropy. This can be done
to any order in the $G_N$ expansion. In particular,  it can be used to compute the quantum
corrections.  The procedure is the following. First we find the smooth bulk solutions for each
integer $n$. The full partition functions around these geometries, including the classical action and
all quantum corrections, gives the $n^{th}$ Renyi entropies.  One then computes the analytic
 continuation in $n$. At order $G_N^0$ this involves computing the one loop determinants around
  each of the classical solutions. There are many difficulties with this method, including
constructing the smooth bulk solutions and then continuing the replica index to
non-integer $n$. Despite these difficulties,   in \cite{Barrella:2013wja}
this method was used to compute the quantum correction in a few cases using
the classical bulk solutions constructed in \cite{Faulkner:2013yia}.
On the other hand,  the formula \nref{bulk-ent} is a shortcut, or an alternative expression,
 for the final answer  in the same way that \nref{TRent} is a shortcut for the
classical version of the replica method.  The final answer \nref{bulk-ent} is
 physically clearer and  easier to compute.

\subsection{Review of the classical argument}

\begin{figure}[h!]
\begin{center}

\vspace{5mm}
\includegraphics[scale=0.65]{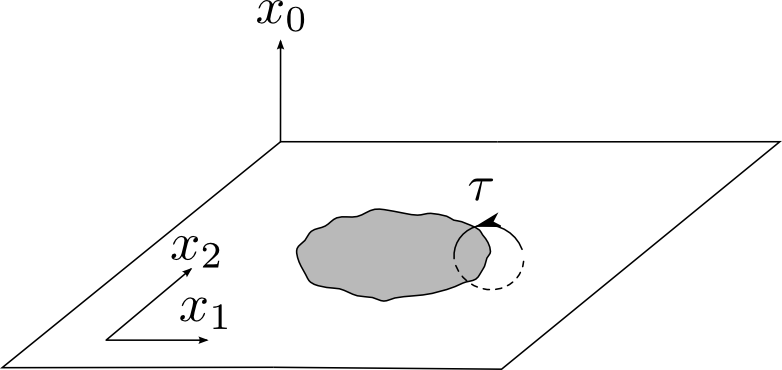}
\vspace{5mm}
\caption{
Slighly deformed disk and angular direction around the boundary.}
\label{tau}
\end{center}
\end{figure}

Let us begin by reviewing the derivation of \nref{TRent} in the classical case \cite{gge}.
First consider the boundary field theory.
The replica method is based on going to euclidean time and then considering an angular direction
with origin at the boundary of region $A$. We label this by $\tau$, with $\tau = \tau + 2 \pi$,  see figure \ref{tau} for an illustration.

We then consider the quantum
field theory in a series of spaces given by the same metric but with
$\tau = \tau + 2 \pi n$, with integer $n$. With the naive boundary metric
this $\tau$ circle shrinks at the boundary of the region $A$. However, we can rescale the metric
by choosing a Weyl factor so that the circle does not shrink according to the boundary metric\footnote{
If the theory is conformal this rescaling does not change the interesting physics. If it is not
conformal we can still do it, but we will have spatially varying dimensionful couplings in the
new space.}. We need
to compute the partition function of the quantum field theory on this sequence of spaces and
then analytically continue in $n$ to compute
\be \la{replicaen}
 S = - \left. \partial_n ( \log Z_n - n \log Z_1 ) \right|_{n=1}  = - Tr[ \rho \log \rho ]
\ee
where $\rho = \rho_A$ is the density matrix of region $A$ in the boundary theory.

 In theories
with gravity duals, the partition functions can be computed by considering    bulk
solutions, $g_n$, which end at the boundary on the geometries we have defined above.
Then one computes the gravitational action and partition functions for these solutions. This
can be done to any order in the $G_N$ expansion. The leading order answer comes from evaluating
the classical action. We discuss this first.

\begin{figure}[h!]
\begin{center}
\vspace{5mm}
\includegraphics[scale=0.8]{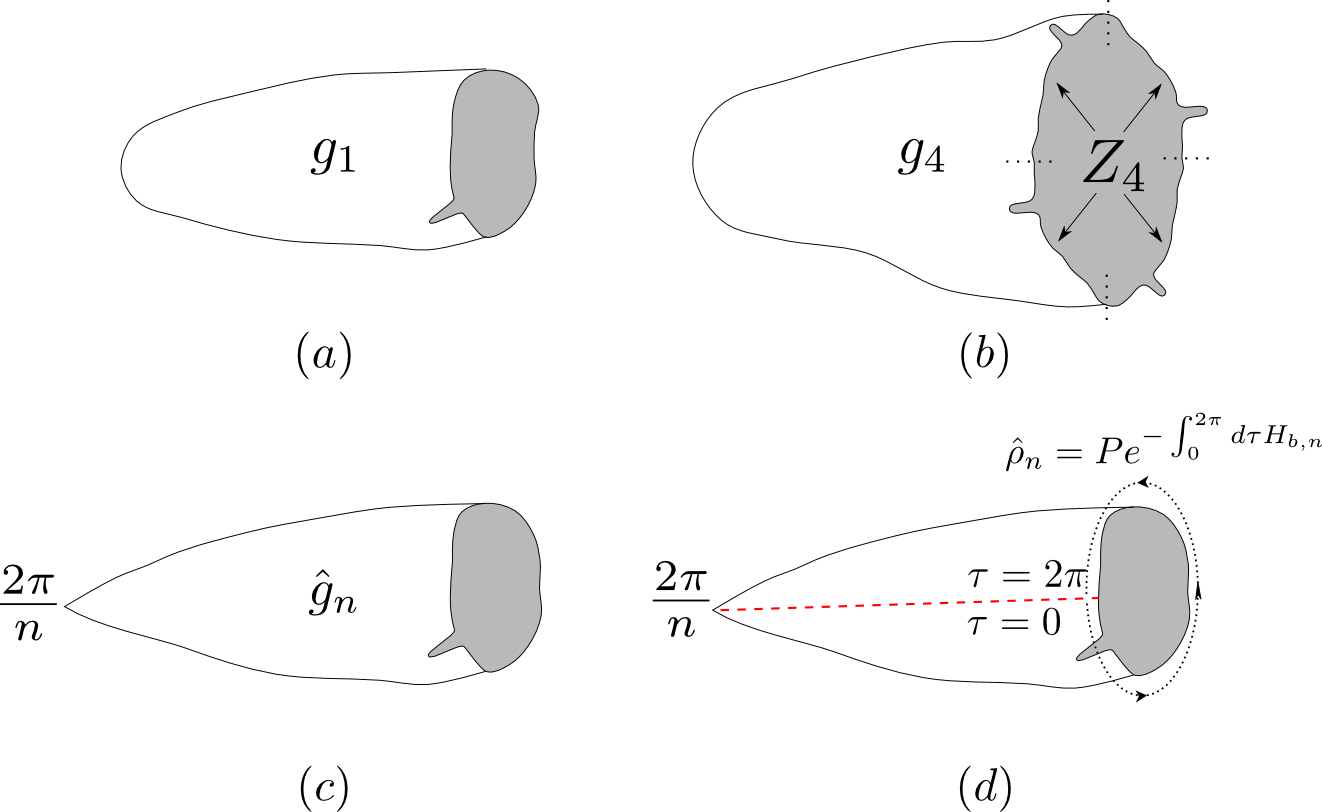}
\vspace{5mm}
\caption{
Computation of the entropy using the replica trick. $a)$ Original geometry with no $U(1)$ symmetry. $b)$ Replicated smooth geometry $g_4$. $c)$ After a $Z_n$ quotient of the $g_n$ geometry of b) we get the
geometry $\hat g_n = g_n/Z_n$. It has a conical singularity with opening angle $2\pi/n$. This geometry
has the same asymptotic boundary conditions as the original one in a). We can analytically continue
this geometry to non-integer values of $n$.
  $d)$ We use the geometries in c) to construct the density matrix $\hat{\rho}_n$.
   $\hat{\rho}_n$ is defined as a path integral on this geometry
with arbitrary boundary conditions at $\tau =0, 2\pi$. It can be computed using the bulk
Hamiltonian for $\tau$ evolution.}
\label{gngeometries}
\end{center}
\end{figure}

 These bulk  geometries, $g_n$,
are typically such that the circle $\tau$ shrinks smoothly in the interior. These
 geometries have a $Z_n$
symmetry generated by  $\tau \to \tau + 2 \pi$, since the metric and all
 other couplings are periodic under this
shift. See figure \ref{gngeometries}.
 It is convenient to introduce the geometries $\hat g_n = g_n/Z_n$. These are bulk geometries
with exactly the same boundary conditions as the original geometry, $g_1$, with $\tau = \tau + 2 \pi$.
However, these geometries typically contain a conical defect, or cosmic ``string'' (a codimension two surface)
with opening angle $2 \pi/n$. These sit at the points where the $Z_n$ symmetry had fixed points,
the points where the circle shrinks.
Then the classical action obeys the condition $I[g_n] = n I[\hat g_n]$. This just
follows from the fact that the classical action is the $\tau$ integral of a local lagrangian
density.  In  evaluating
$I[\hat g_n ]$ we {\it do not } include any contributions from the singularity, not
even a Gibbons-Hawking boundary term near the singularity\footnote{
We still include the Gibbons-Hawking  boundary
term at the $AdS$ boundary, as usual.}.  We simply integrate the
usual bulk  lagrangian away from the singularity. We can now analytically continue the geometries
$\hat g_n$ to non-integer $n$. They have the same boundary as the $n=1$ solution, but in the
interior they contain cosmic ``string''  singularity of opening angle $2\pi/n$.
When $n\to 1$ we have a very light cosmic string. The minimal area
condition comes from the equations of motion of this cosmic string  and the area formula
\nref{TRent} follows essentially from its action, see \cite{gge} for more details.

\subsection{Quantum argument}

This is a generalization of the black hole discussion  in
\cite{Solodukhin.BH,offvson}  to situations  without the $U(1)$ symmetry.

At the quantum level,
the replica trick instructs us to compute the partition function of all the bulk quantum
 fields around the
black hole geometry. This involves computing the functional determinants for the quadratic
fluctuations around the geometries $g_n$ \footnote{Part of the bulk fields could be strongly
coupled.  For example, we can have a non-trivial CFT in the bulk. In that case,
 the bulk computation is
more complicated, but the principle is the same (at this order in the $G_N$ expansion):
 computing the partition function in the
geometry $g_n$.}. In performing this computation we can view $\tau$ as a time evolution, so that
the quantum partition function can be written as
\bea \label{qu-trace}
Z_{q,n} &=& Tr[ P e^{ -\int_0^{ 2 \pi n } d\tau H_{b,n}(\tau)} ] = Tr[ \hat \rho_n^n ] ~,~~~~~~
\cr
\hat \rho_n & \equiv &  P e^{ - \int_0^{2 \pi } H_{b,n}(\tau) }   \label{rhohatdef}
\eea
Here $H_{b,n}(\tau)$ is the bulk time dependent hamiltonian that evolves the system along the $\tau$
direction\footnote{$H_{b,n}(\tau)$ is a local integral over a constant $\tau$ spatial slice.   This should not be confused with the so called  ``modular hamiltonian'', $K$, defined through $e^{-K}=P e^{ - \int_0^{2 \pi } H_{b,n}(\tau) } $ which is a non-local operator.
 }. It depends on $n$ because the equal $\tau$ slices of the geometry $g_n$ do depend on $n$.
 In the second equality we have used the fact that $H_b(\tau) = H_b(\tau + 2 \pi )$.
We have also defined a bulk (non-normalized) density matrix $\hat \rho_n$. The $n$ subscript
reminds us that the definition depends on $n$ because the bulk geometry depends on $n$.
In fact, we can assign $\hat \rho_n$ also to the bulk geometry $\hat g_n = g_n/Z_n$.
Up to now the discussion  was for integer $n$.

Now we analytically continue to non-integer $n$ as follows.
We consider the bulk geometry $\hat g_n$ that we defined for the classical computation.
We define again $\hat \rho_n$ as given by the same expression as in \nref{rhohatdef}. Now $H_{n,b}$ is
a Hamiltonian defined on equal $\tau$ slices of the geometry $\hat g_n$, with non-integer $n$.
  In summary, we define the partition function for non-integer $n$ via
\be \la{partnonint}
Z_{q,n} = Tr[ \hat \rho_n^n]  ~,~~~~~~~~~~~~ \hat \rho_n  \equiv   P e^{ - \int_0^{2 \pi } H_{b,n}(\tau) }
\ee
Here we are ignoring  UV  divergencies. More precisely, we can consider a
UV regulator that is local and general covariant so that the discussion is valid for the regulated theory.

We can now write the expression for the quantum correction as
\bea
S_q  &=&- \partial_n \left(  \log Z_{q,n} - n \log Z_{q,1} \right)_{n=1} =
-\partial_n( \log  Tr[ \hat \rho_n^n] - n \log Tr[\hat \rho_1]  ) _{n=1} =
\cr
 &=&
S_{\rm bulk-ent} + S_{\cdots}
\cr
S_{\rm bulk-ent} &=& - \partial_n ( \log Tr[  \rho_1^n ]  - n \log Tr[ \rho_1 ] )_{n=1} ~,~~~~~~~~~~~
S_{\cdots} \equiv -  {
Tr[ \partial_n \hat  \rho_n ] |_{n=1} \over Tr[ \rho_1 ] }
\eea
Here $S_{\rm bulk-ent}$  involves only $\rho_1 \equiv \hat \rho_1$,  which is the
density matrix in the original ($n=1$) geometry. This term computes the bulk entanglement
entropy. The second term, $S_{\cdots}$,
 arises due to the  $n$ dependence of the bulk solution and gives rise to the dots in
\nref{bulk-ent}. Let us  find a more explicity expression for this term.
  For simplicity,   we
 assume that the solution is such that only the metric is non-zero in the classical background
and the rest of the fields are zero. This can be easily generalized.
To evaluate $S_{\cdots}$ we go again to the Lagrangian formalism.
The Lagragian, ${\cal L}(\hat g_n , h, \varphi)$,
depends on the background metric and the small fluctuations of all the fields: the metric
fluctuations, $h$, as well as all the other fields denoted by $\varphi$.
We can then write
\be \la{Dots}
S_{\cdots}  = \langle  \int d\tau \partial_n {\cal L } \rangle =
\int d\tau  \langle E_{\mu \nu}(\hat g + h,\varphi) \partial_n \hat g^{\mu \nu } + d \Theta( \hat g , h,\varphi ; \partial_n \hat g ) \rangle  - \int d\tau d\Theta( \hat g , \partial_n g )
\ee
where the brackets indicate quantum expectation values. In other words, we integrate over
the fields $h$ and $\varphi$.
Here $E_{\mu \nu}$ represent equations of motion for the metric.
 These do not vanish because the quantum fluctuations are off shell.
  And $\Theta$ is related to all the partial integrations involved in going  from a
variation of the lagrangian to the equations of motion. We are using a notation similar to \cite{Iyer:1994ys}, where
the reader can find  explicit expressions. $\Theta$ is linear in $\partial_n \hat g $.
The $\Theta$  term is the same as the one that   gives rise to the Wald-like entropy formula \cite{Iyer:1994ys}. We say Wald-like because we are considering a situation without a $U(1)$ symmetry.
For the usual two derivative action, it gives rise to the area formula. Here we are evaluating it
for a generic off shell configuration (since we have  general variations $h, ~\varphi$) and computing the
expectation value. We have also subtracted the classical result.
The simplest example where this term is nonzero is the following.
 Consider  a theory with a scalar field with a coupling
$\zeta \phi^2 R$. If the scalar field is zero in the classical solution, this does not contribute to
the classical black hole entropy. However, if we consider the small fluctuations of $\phi$, we will
get a term proportional to  $ \zeta \langle \phi^2 \rangle ( {\rm Area} )$. Such a term arises from the
$\Theta$  term in \nref{Dots}. In general,  we denote such terms as $ \langle \Delta  S_{\rm W-like} \rangle$ \footnote{  Note that for solutions where $R=0$, we do not expect any $\zeta$ dependence on $S_q$ or $S_{bulk-ent}$. $S_{bulk-ent}$ does not depend on $\zeta$ and one can easily show that the $\zeta$ dependence on the finite part of $S_{...}$ cancels between $\delta A$ and $ \langle \Delta  S_{\rm W-like} \rangle$ terms.} . This is
the expectation value of the formal expression for the Wald-like entropy\footnote{
In situations without a
$U(1)$ symmetry, the general Wald-like expression for a
general higher derivative theory is not known. For the purposes of this discussion we simply assume that
such an expression exists. In the
case of an action with  $R^2$  terms  the expression was found in \cite{highermyers,higherparnachev,Bhattacharyya:2013jma,squashedcones}.
}. We expect that the graviton
gives rise to possible contributions to this term.

\begin{figure}[h!]
\begin{center}
\vspace{5mm}
\includegraphics[scale=0.5]{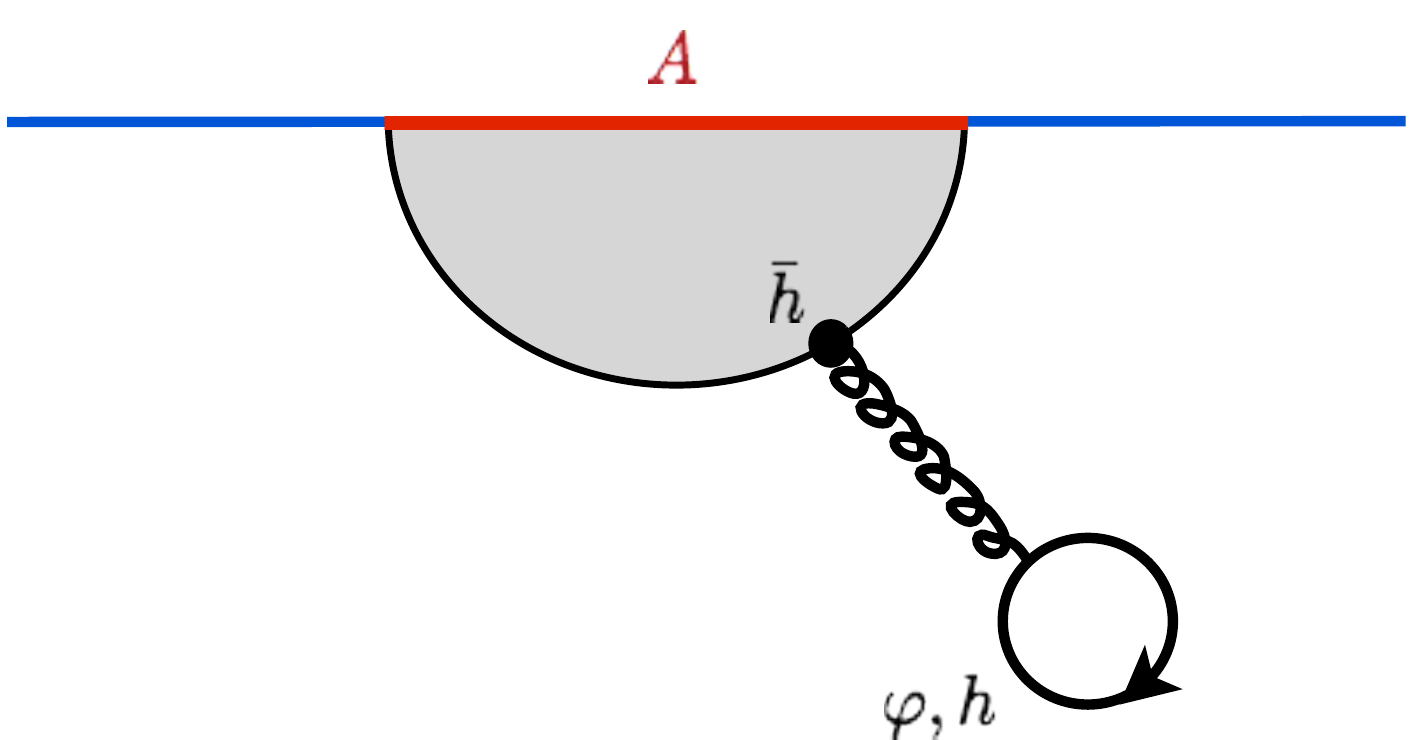}
\vspace{5mm}
\caption{\label{figdA}
The contribution to $S_{\cdots}$ from the change in the area of the minimal surface, $\delta A$,
 due to the quantum corrections of the background. We can interpret this diagram
as solving \eqref{backh} for $\bar{h}$ in terms of 1-loop stress tensor. We need
to solve for $\bar{h}$ along the minimal surface and integrate the stress tensor
over all space.
 }
\end{center}
\end{figure}

Now let us focus on the first term in \nref{Dots}. The equations of motion are non-zero because
we are considering quantum fluctuations. We can formally write this term as
\be \la{expecval}
 \int d\tau \langle    E _{\mu \nu} \rangle   \partial_n \hat g^{\mu \nu}   = -\frac{1}{2} \int d\tau \langle    T _{\mu \nu} \rangle   \partial_n \hat g^{\mu \nu}
\ee
Here we have viewed the quantum expectation value of the equations of motion as a quantum generated
expectation value for the stress tensor.
This expectation value of the equations of motion   will force us to   change
in the classical background. Indeed, to avoid ``tadpoles'' we will need to change the classical background
$\hat g \to \hat g + \bar h$, where $\bar h$ is small classical correction of order $G_N$ in such a way that
\be
\label{backh}
E_{\mu\nu}(\hat g + \bar h)  = -  \langle    E _{\mu \nu} \rangle  = \frac{1}{2} \langle    T _{\mu \nu} \rangle
\ee
where we are expanding the left hand side only to first order in $\bar h$.
We can then reexpress \nref{expecval} as
\be
\la{fina}
\int d\tau E(\hat g + \bar h)  \partial_n \hat g  =
 \partial_n I_n(\hat g_n+\bar h )|_{n=1} - \int d\tau d \Theta(\hat g + \bar h, \partial_n \hat g )
\ee
To first order in $\bar h$,   $I_n(\hat g_n + \bar h) = I_n(\hat g_n)$
due to the equations of motion for
$\hat g$. In \nref{fina} we are considering $n$ very close to one.
 Here $\bar h$ is the solution for $n=1$, and we have kept it fixed as we vary $n$ away from one.
We have also ignored higher order terms in $\bar h$.
The right hand side of \nref{fina} can be then
 rewritten as the change in the area due to the shift of the classical solution,
$ {\delta A \over 4 G_N }$. Since the change in the background is of order $G_N$, this term is of order
one. In a general higher derivative theory this will presumably become $\delta S_{\rm W-like}$.
A diagrammatic interpretation of this contribution is given in figure~\ref{figdA}.

In addition, we should add terms arising from the counterterms that render the bulk quantum
theory finite.
 Such counterterms are given by local expressions in terms of the metric and the
curvature, etc. Thus they look like the classical action itself. They contribute to the entropy
via local terms of the same form as the ones we get for a general higher derivative local action.
For example a counterterm of the form
$ { 1 \over \epsilon^{D-2} } \int R $ gives a contribution  $ { ({\rm Area} )\over \epsilon^{D-2} } $. There are
similar contributions from higher derivative terms.
 We just apply the Wald-like formula for the counterterms\footnote{As we mentioned this
formula is unknown for general non-U(1) invariant situations. However see  \cite{highermyers,higherparnachev,Bhattacharyya:2013jma,squashedcones}
 for $R^2$ corrections.
Here we simply assume that such a formula exists.} .

In conclusion,
 the full expression for the quantum correction to the entropy is given by
\be \la{finalqu}
S_q = S_{bulk-ent} + { \delta A \over 4 G_N} +  \langle \Delta  S_{\rm W-like} \rangle  + S_{\rm counterterms}
\ee
The first term is the bulk entanglement. The second is the change in the area due to the shift in the
classical background due to quantum corrections. The third is the quantum expectation value
of the formal expression of the Wald-like entropy. The final term arises because we need to introduce
counterterms in order to render the computation finite\footnote{Some aspects of these counterterms
have been discussed recently in \cite{Cooperman:2013iqr}.}.
 The last three terms in \nref{finalqu}  fill in the dots in \nref{bulk-ent}. Some articles, e.g. \cite{Solodukhin.BH}, compute the entanglement entropy by smoothing the tip of the cone and,
 when fields are coupled to curvature, they obtain an extra contact term,
 this is precisely our Wald-like term, $S_{\rm reg-cone}=S_{bulk-ent}+\langle \Delta  S_{\rm W-like} \rangle$.

Let us finish with some comments. The expression \nref{finalqu} for the case of black holes
was discussed in \cite{Solodukhin.BH,offvson}\footnote{
In the black hole case, where one has a $U(1)$ symmetry, it is easier to define the quantum
computation for non-integer $n$. Here we had to define it as \nref{partnonint}.} .  Notice that, in the black hole case, we can  compute the entropy using the
Gibbons-Hawking method, which is to change the period of $\tau$  (called $\beta$), considering
always the smooth solution. In this case,
 we get the full quantum result  from the determinants, computed
on the $n$- (or $\beta$-)dependent geometry.
In other words, at this order,
there is no need to shift the classical
background due to quantum corrections, or to evaluate quantum expectation values of the
formal expression for the Wald entropy. \footnote{  For example, this has
been carried out explicitly to find the  logarithmic corrections to black hole entropy, see \cite{Sen:2012dw} and references within. }
However,  if we evaluate the quantum correction using
bulk entanglement (as opposed to the Gibbons-Hawking
method) we need to take them into account to get the right answer. Similarly, if we compute the quantum correction using the replica trick, we can just compute the
determinants, and analytically continue them without worrying about the changes in the classical
background due to the quantum corrections, as  was done for $AdS_3$ in \cite{Barrella:2013wja}.

The last three terms in \nref{finalqu} are given by local integrals on the original minimal surface.
Thus, they contribute terms which are qualitatively  similar to the classical contribution.
The classical Ryu-Takayanagi formula was shown to obey various nontrivial
 inequalities also obeyed by entanglement entropy \cite{Headrick:2007km}. One of these is the strong subadditivity
condition. In fact, this inequality follows from the fact  that we are
minimizing a quantity in the bulk \cite{Headrick:2007km}.
Thus if we add the last three terms in \nref{finalqu} to the Ryu-Takayanagi formula, we still get
a result that can be viewed as the minimization of a local expression. To order $G_N^0$,
the corrections in
\nref{finalqu} do not change the shape of the surface because they are small corrections.
Moreover, the bulk entanglement contribution, the first term in \nref{finalqu}, obeys the entropy
strong
subadditivity condition on its own, since it can be viewed as a field theory computation in the bulk.
Thus, we have argued that the classical plus first quantum contribution
 should also obey the strong subadditivity condition.

\section{Applications}

Here we discuss some applications of the above formula.
We will concentrate on cases where the quantum correction gives a
qualitatively new effect.

\subsection{ Almost gapped large $N$ theory}

Consider the Klebanov-Strassler  theory in the large $N$ limit, where it is
described by the gravity dual found in \cite{ks}.
The shape of the corresponding geometry is such that most of the bulk fields give rise to
massive excitations from the four dimensional point of view.
 The only massless excitations are associated
to the spontaneous breaking of the $U(1)$ baryon symmetry \cite{Aharony:2000pp,Gubser:2004qj}. Since it is a supersymmetric
theory, the usual Goldstone boson is part of a  massless chiral superfield.

\begin{figure}[h!]
\begin{center}
\vspace{5mm}
\includegraphics[scale=.75]{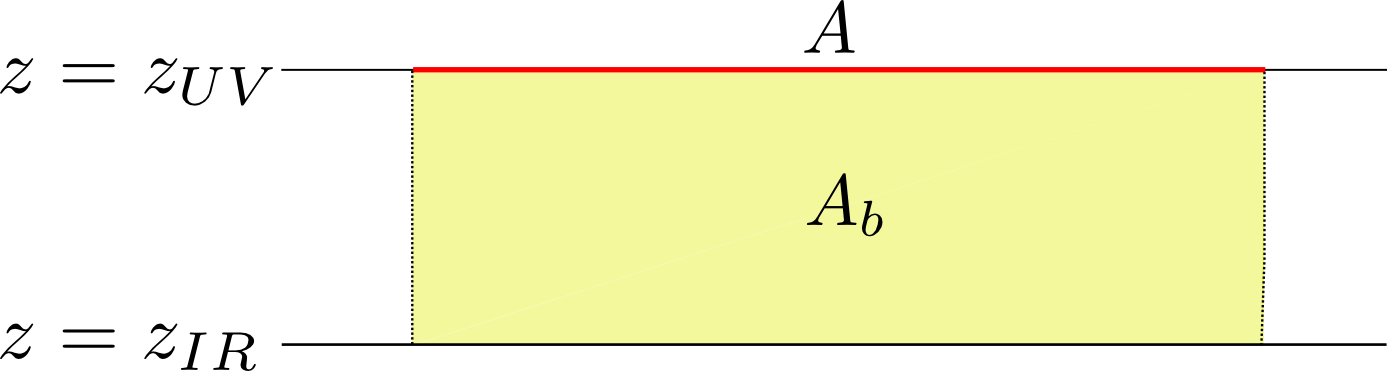}
\vspace{5mm}
\caption{Shape of the minimal area surface in the Klebanov-Strassler theory.  The yellow region is the interior. The quantum
correction is given by the entanglement between the interior and the exterior.   }
\label{KSquantum}
\end{center}
\end{figure}

Now consider a region $A$ of a size which is larger than the inverse mass of the lightest massive modes.
The classical contribution for such a region was computed in \cite{ent.and.confinement}.
This arises from a minimal area surface which comes down from the boundary into the bottom
of the throat with a topology as indicated in figure \ref{KSquantum} .
The result is that it goes as
\be \la{ConfTh}
 S_{cl}  \propto   { c_0 R^2   +  {\rm constant } }  + \cdots
\ee
for large $R$, where $R$ is the size of the region. Here $c_0$ has both UV divergent and finite contributions.  $c_0$ is proportional to $N^2$ \footnote{ Here by $N$ we mean the value of $N$ in
the last step of the cascade \cite{ks}.
The UV divergent contribution has a larger effective value of $N$. }

The quantum correction  is given by
 the entanglement in the bulk between the interior and the
exterior of region bounded by the Ryu-Takayanagi minimal area surface in the bulk, see figure \ref{KSquantum}.
For a large region, we can approximately compute the bulk contributions by doing a Kaluza-Klein
decomposition of all the bulk fields, and then doing the entanglement computation in four dimensions.
 To the order we are working,  all the bulk fields are free.
All the massive bulk modes contribute only with terms that give rise to contributions similar to
\nref{ConfTh}. However, the massless modes (two bosons and two fermions) give rise to a  qualitatively new
logarithmic term of the form
\be
S_{\rm q - log} = - \alpha \log R \Lambda
\ee
where $\Lambda$ is the scale setting the mass of the massive modes.
Here
$\alpha$   is a numerical constant that depends on the shape
of the region \cite{Solodukhin:2008dh}.
For a spherical region $\alpha = 4 a$ where $a$ is the conformal anomaly coefficient
for a chiral superfield, $a= - { 1 \over 48}$.

A similar correction to the Ryu-Takayanagi
 formula was argued for in \cite{Fujita:2009kw}. In Section 3 of \cite{Fujita:2009kw}
  they   consider an AdS soliton geometry which is  dual to a 3d confining gauge theory. A Chern-Simons term was added to the boundary theory resulting in a topological
theory in the IR. The expected
topological term in the entanglement entropy is reproduced by the  entanglement
of bulk fields. This provides a further check of
\eqref{bulk-ent}.

\subsection{Thermal systems in the bulk}
\label{ThermalStates}

\begin{figure}[h!]
\begin{center}
\vspace{5mm}
\includegraphics[scale=.75]{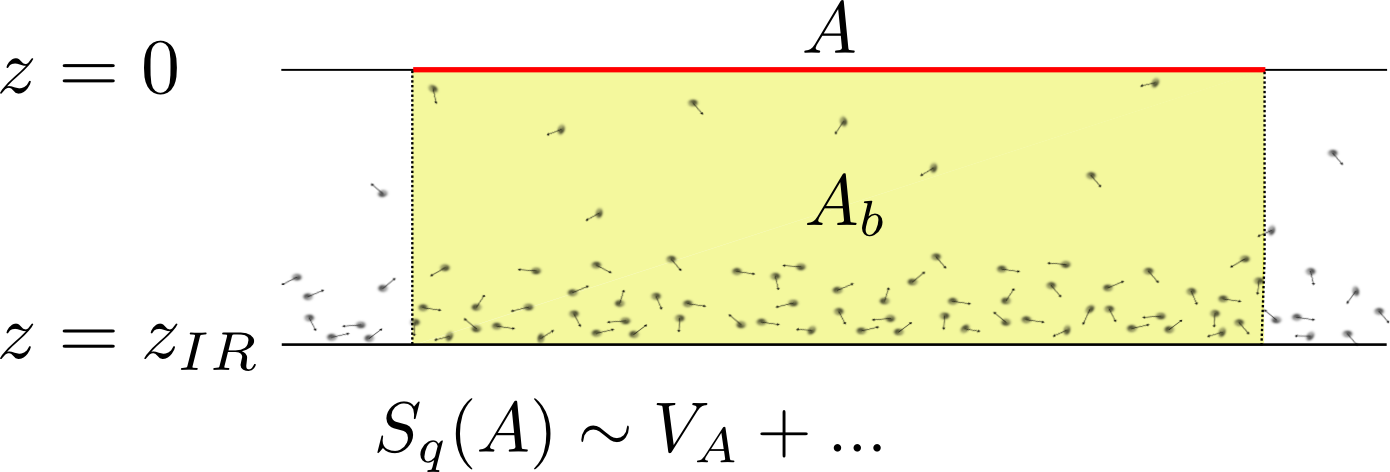}
\vspace{5mm}
\caption{   Confining theory and thermal gas in the bulk.  Here $V_A$ is the volume of region $A$ in the
boundary}
\label{ThermalGas}
\end{center}
\end{figure}

We can consider a confining theory whose geometry can be modelled by  an AdS space with
an infrared end of the world brane.  In this case, let us consider a theory with no massless modes.
Then the entanglement entropy of  a large region of size $R$ will behave as in  \nref{ConfTh}.
This will be the case as long as we consider the theory in the vacuum.
However, if we consider the theory in a thermal   bulk state, with a gas of particles in the bulk,
we get a contribution to the entropy from this gas. We are considering the phase with no black brane.
 Then we get  a contribution proportional to the
volume, $S(A) \propto V_A$, in addition to \nref{ConfTh}. This contribution is of order $G_N^0$ (or $N^0$). We obtain this contribution from the bulk entanglement entropy of region $A_b$,
 eee figure \nref{ThermalGas}.

Another case which is qualitatively similar arises when we consider a fermi surface in the bulk\footnote{We thank S. Hartnol for pointing out
this application.}. 
Since we end up computing the bulk entanglement entropy, we reproduce the logarithmic terms that 
are expected in that context \cite{Wolf:2006zzb,Gioev}. 
This is important for applications of AdS/CFT to non-Fermi liquids. See for example
\cite{Faulkner:2009wj,Hartnoll:2010gu,Sachdev:2011ze}, where such logarithmic
 violations are expected due to the appearance of bulk fermi surfaces. This should be contrasted
with \cite{Ogawa:2011bz} where the logarithmic violations to the entanglement entropy where found from the leading geometric term.

 \subsection{ Non-contractible circle}

If the $\tau $ circle that appeared in our discussion in section \ref{Argument}
 is not contractible in the bulk,
then the classical contribution to the entropy vanishes. In this case, the whole contribution
to the generalized gravitational entropy comes from the quantum correction. It involves the propagation
of the quantum particles around the bulk circle. It is a finite contribution. In the case that the system
has a $U(1)$ symmetry, this is just the thermal entropy of a gas of particles in the bulk. In  general,
this setup leads to a bulk mixed state under analytical continuation to Lorentzian signature and we just
get the entropy of this bulk mixed state.

\subsection{Mutual information, generalities  }
\label{MIGeneralities}

\begin{figure}[h!]
\begin{center}
\vspace{10.5mm}
\includegraphics[scale=0.85]{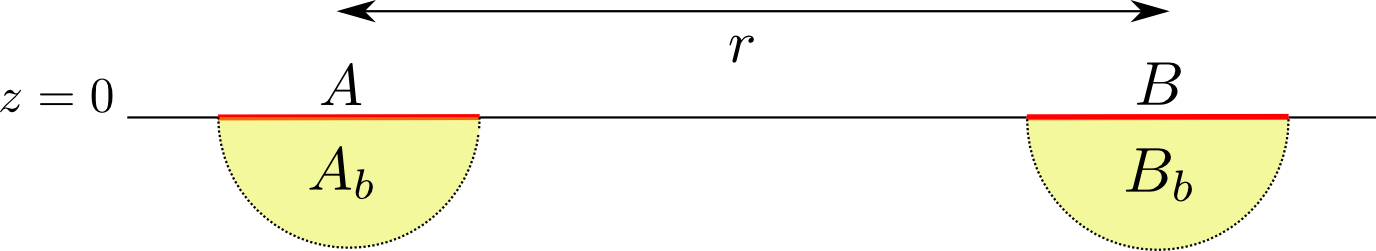}
\vspace{5mm}
\caption{  We consider two regions $A$ and $B$ on the boundary which are separted by a long
distance, $r \gg r_A,~r_B$, where $r_{A,B}$ are their sizes. The minimal area surfaces have the shape
indicated. In the bulk, they define regions $A_b$ and $B_b$, which are shown in yellow.
 The surface for $S(A \cup B)$  is simply
the sum of the two surfaces.  }
\label{ClassicalMutual}
\end{center}
\end{figure}

For two disjoint regions, $A$ and $B$, we define the mutual information
\be
I(A,B) =
S(A) + S(B) - S(A  \cup  B )
\ee
A   feature of the  Ryu-Takayanagi formula is that, for well separated disjoint regions,
the mutual information is zero \cite{Headrick:2010zt}. See figure \ref{ClassicalMutual}.
In other words, the classical bulk answer is zero. This is due to the fact that the surface for
$S(A\cup B)$ is the union of the surfaces that we use to compute   $S(A)$ and  $ S(B)$.
 We will see that the
quantum correction gives us something different from zero.
Note that  all the local contributions (coming from the second, third and fourth terms in \nref{finalqu}) also  cancel for the same reason as in the classical case.  Thus
 mutual information comes purely from the bulk entanglement term (the first term in \nref{finalqu}).
Thus the quantum contribution to the mutual information is simply equal to the bulk
mutual information for the two bulk regions:
 \be
 \la{BulkEnt}
 I(A,B)=I_{bulk,ent}(A_b,B_b)
 \ee
  Here $A$ and $B$ are
two regions in the boundary CFT. $A_b$ and $B_b$ are the two corresponding regions in the bulk,
see figure \ref{ClassicalMutual}.
As explained in
\cite{raams1,raams2}, a non zero answer is necessary for having non-vanishing correlators.
The argument is based on the general bound for correlators
\cite{ciracbound}
\be \la{boundcorr}
I(A,B) \ge \frac{(\langle {\cal O}_A {\cal O}_B \rangle-\langle {\cal O}_A \rangle \langle  {\cal O}_B \rangle )^2}{2 |{\cal O}_A |^2 |{\cal O}_B |^2 }
\ee
where $|{\cal O}_A|$ is the absolute value of the maximum eigenvalue\footnote{Of course, we should
choose ${\cal O}_A$ to be a suitably smeared function of a local operator so that the maximum
eigenvalue is finite. For example,   ${\cal O}_A \sim e^{ i \int  O(x) g(x) }$, where $g(x)$ is a localized smooth function.}.
Thus, the non-zero one loop correction will enable us to obey this bound. We will discuss this in
more detail below.

\subsubsection{  Long distance expansion for  the mutual information in quantum field theory }

Here we consider two disjoint regions, $A$ and $B$ that are separated by a large distance in
the boundary theory. In this situation, one can do a kind of operator product expansion for the
mutual information. As discussed in \cite{Casini:2008wt,Headrick:2010zt,Calabrese:2010he,Cardy:2013nua},
the expected leading contribution comes from the exchange of a pair of operators
each with dimension $\Delta$ \footnote{An idea for an OPE expansion of mutual information was discussed in \cite{wilsonloopMI}. However, we think that it is not correct because it includes the exchange of
 single particle states, as opposed to two particle states.  }. In other words, we have \cite{Casini:2008wt,Headrick:2010zt,Calabrese:2010he,Cardy:2013nua}
\be \label{MutualOPE}
I(A,B) \sim \sum C_\Delta { 1 \over r^{4 \Delta} }+ \cdots
\ee
where $C_\Delta$ comes from squares of OPE coefficients. These OPE coefficients $C^A_\mathcal{O}$  arise by replacing  region $A$ of the replica space by a sum $ \sum C^A_\mathcal{O}  \mathcal{O}$  over local operators in the $n$ copies of  the original CFT.  Such operators take
the form of products of operators of the original CFT living on the different replicas.
  Once
we have have these OPE coefficients   we can find:
\be
\label{coefficient}
C_\Delta =  \partial_n\left[ \sum C^A_\mathcal{O} C^B_\mathcal{O} \right]_{n=1}
\ee
where the sum is over all operators contributing at the same order as \eqref{MutualOPE}. This involves
sums over operators in different replicas and the analytic continuation in $n$ appears non-trivial.

\begin{figure}[h!]
\begin{center}
\vspace{10mm}
\includegraphics[scale=1]{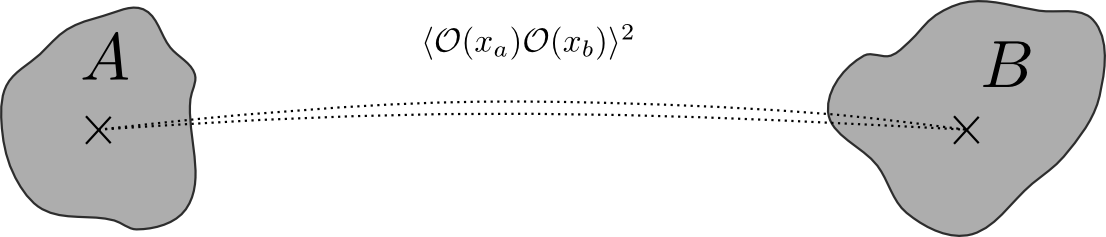}
\vspace{5mm}
\caption{   OPE-like expansion for mutual information.}
\label{OPEMutual}
\end{center}
\end{figure}

For a single operator living on a single replica the OPE coefficient $C_{\cal O}^A$,
 in principle, could
be calculated. However,  it vanishes as $ (n- 1)$ since the one point
functions of the un-replicated space vanishes. Therefore,  the square of the OPE
coefficient in \eqref{coefficient}   vanishes at $n=1$.
 The two operator case in \eqref{MutualOPE} gives
the first non-zero answer.  We expect that the leading contribution comes from pairs of operators
with lowest anomalous dimension.

At integer $n$ we are doing a standard OPE expansion in terms of operators of the replicated theory.
 However, the final result at $n=1$ cannot be interpreted as an ordinary OPE expansion in the
 original theory. For example, the leading behavior in \nref{MutualOPE} might not be reproduced by
 operators of the original theory. For example, the theory, at $n=1$, might not have an operator
with dimension $\Delta' = 2 \Delta$ to reproduce \nref{MutualOPE}\footnote{ For example, in the Ising model, the leading term
 comes from the spin operator of dimension $\Delta =1/4$. However there is no (spin zero) operator
 in the theory with dimension $1/2$ that can reproduce \nref{MutualOPE}.}.
   In general, the individual OPE coefficients cannot be continued
to $n=1$.  However the sums of squares of all the OPE coefficients contributing
at the same order in \eqref{MutualOPE} can be continued to $n=1$ \cite{Calabrese:2010he}.
Here we will not compute the OPE coefficients, we simply focus on the $r$ dependence.

Notice that this behavior of the mutual information, \nref{MutualOPE},
 is consistent with the bound \nref{boundcorr}. In addition, this implies that the $C_\Delta$
 coefficient for
the lightest operator cannot vanish.

In large $N$ theories, the standard large $N$ counting rules imply that the
OPE coefficients $C^A_{\cal O}$ for the leading contribution are of order one, since they
come from a connected two point function in the replicated geometry. This is in the normalization where the two point function of single trace operators is normalized to one. Thus, the leading
contribution to $C_\Delta$ vanishes at order $N^2$ and is non-vanishing at order one.
Similar large $N$ counting for more general operators  leads us to expect that the mutual information vanishes exactly at order $N^2$ in large $N$ theories,
for well separated regions,
  as is the case in large $N$ theories with gravity duals.
   For this argument, the crucial feature is
  that the contribution from the exchange of a single operator vanishes\footnote{This is no longer true
  for the mutual {\rm Renyi} entropies \cite{Headrick:2010zt}.}.

We can similarly consider mutual information in non-conformal theories. For example we can
consider a massive theory. In this case the long distance expansion can be done in terms of the
excitations of the massive theory, in terms of the lightest massive excitation. As before, these
excitations will propagate along the $n$ separate copies of the replicated theory. And the leading
contribution comes from pairs of the lightest particle\footnote{
In general, the contribution from the exchange of a single particle should vanish when $n\to 1$.
In free theories,
the single particle contribution vanishes for all $n$ due to a $Z_2$ symmetry that multiplies the
field by a minus sign.}. Again the bound \nref{boundcorr} implies
that the corresponding coefficient cannot vanish.
So far we have discussed theories in flat space. We can similarly consider theories in curved
spaces. Again, for well separated regions, we have a long distance expansion of the mutual
information that involves the propagation of the lightest excitations, but now in curved spacetimes.
Thus the mutual information behaves as
\be  \label{MIcurved}
I(A,B) \sim C  G(x_A,x_B)^2 + \cdots
\ee
where $G$ is the propagator for the lightest excitation of the theory in the curved manifold. More
precisely, the one whose $G(x_A,x_B)$ propagator is the largest.

\subsubsection{ Long distance expansion for mutual information using gravity duals}

Now we consider a theory with a  gravity dual. For
well separated regions,  as argued around \nref{BulkEnt},  the leading order term comes from the bulk entanglement between regions
$A_b$ and $B_b$, see figure  \ref{ClassicalMutual}.
In this approximation, we have a quantum field theory in a fixed background geometry. Then the
long distance expansion of the mutual information reduces to the expression in \nref{MIcurved}, where
we should consider the lightest bulk particle. If the theory reduces to pure gravity in the bulk, then
this is the graviton.  Again, the coefficient is non-zero due to the bulk version of \nref{boundcorr}.

But at long distances $G(x_{A_b},x_{B_b}) \sim { 1 \over |x_A - x_B|^{ 2 \Delta } } $ due to the
standard AdS/CFT dictionary \cite{Gubser:1998bc,Witten:1998qj}. Here
$x_{A_b}$ is some point in the bulk region $A_b$ and $x_A$ is some point in the
boundary region $A$.   Inserting this into \nref{MIcurved} we reproduced the expected field theory result \nref{MutualOPE}.

\subsection{Corrections to the Entanglement Plateaux}

Another situation where we expect quantum corrections to be the dominant
answer comes from considering entanglement entropy for subsystems in thermal states.
They satisfy the Araki-Lieb inequality \cite{arakilieb}:
\be
\label{al}
\Delta S =  S(\rho) - \left| S_{A^c} - S_{A}  \right| \geq 0
\ee
where $\rho$ is the density matrix describing the state of the full system. Here $A^c$ is the
complement of region $A$ in the boundary theory ($A \cup A^c$ gives the full system).
For a thermal state $S(\rho)$ is just the thermal entropy of the full system.

\begin{figure}[h!]
\begin{center}
\vspace{5mm}
\includegraphics[scale=.55]{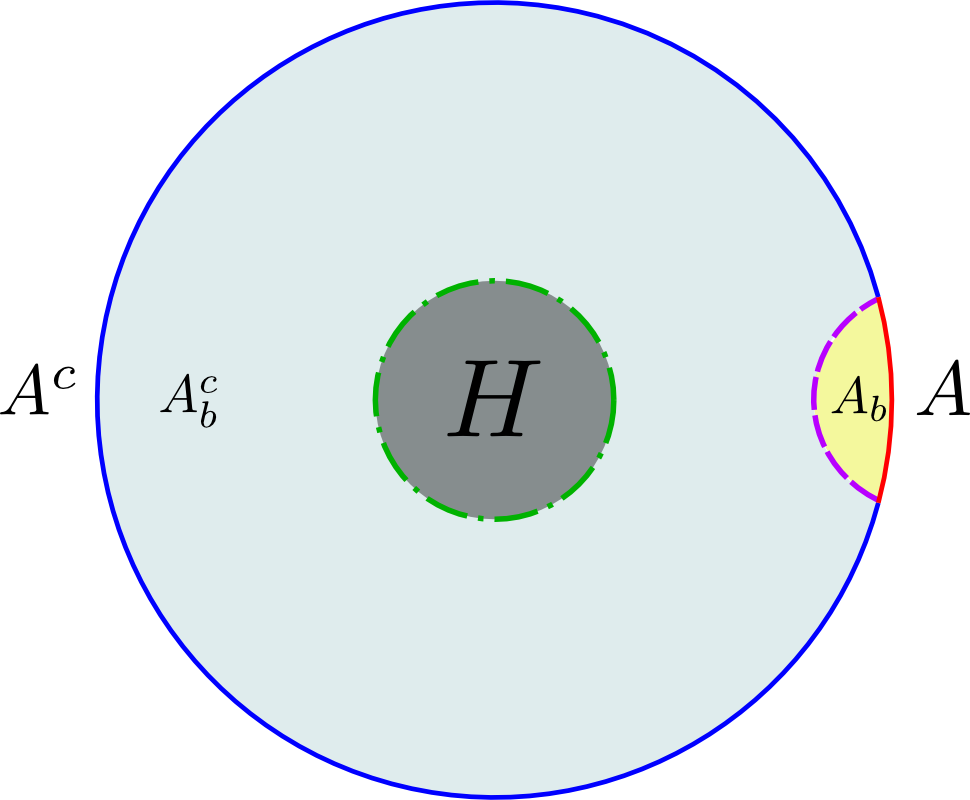}
\vspace{5mm}
\caption{ We consider a small region $A$ and its complement $A^c$ in a finite temperature state.
The bulk contains a black hole. The region $A_b$ is the region outside the black hole horizon.
The minimal surface that gives the leading anwer to $S(A)$ is the the one indicated by a purple dashed line
surrounding region $A_b$. The surface associated to $S(A^c)$ is the one associated to $S(A)$ plus
the black hole horizon. The thermal entropy is computed by the surface at the black hole horizon.
The region $H$ is  the interior of the black hole.   }
\label{plateaux}
\end{center}
\end{figure}

In holographic theories this inequality can be saturated when $A$ is small enough
 (or equivalently $A^c$ is small.) This was discussed
extensively in \cite{Hubeny:2013gta} where this saturation was named the Entanglement Plateaux (see also \cite{Blanco:2013joa,RT,Headrick:2007km}.)
That is, for region $A$ small enough the minimal surface for region $A^c$ is the disconnected
sum of the minimal surface for region $A$ and the horizon of a black hole in the bulk,
see figure \ref{plateaux}.  The thermal entropy is computed by the black hole horizon.
Thus the classical answer gives a vanishing contribution to \nref{al}.
 In the bulk, the first non-zero
 contribution to \nref{al} comes from the bulk entanglement contribution to
the quantum correction \nref{finalqu}.   This reduces to
\be \label{Platin}
\Delta S = S_H - S_{A^c_b} + S_{A_b} = S_H  + S_{A_b} - S_{H \cup A_b } = I(H,A_b) >  0
\ee
where region $H$ is the region behind the horizon. We are imagining we have the eternal black hole and
region $H$ is the second bulk space joined to the first by the Einstein-Rosen bridge.
 We see that $\Delta S$ is the same as the bulk mutual information of regions $H$ and
$A_b$. This is positive by the subadditivity condition applied to the bulk field theory. The
inequality in  \nref{Platin} is strict because of \nref{boundcorr} applied to the bulk theory.

\subsection{ EPR pair in the bulk }

\begin{figure}[h!]
\begin{center}

\vspace{5mm}
\includegraphics[scale=1]{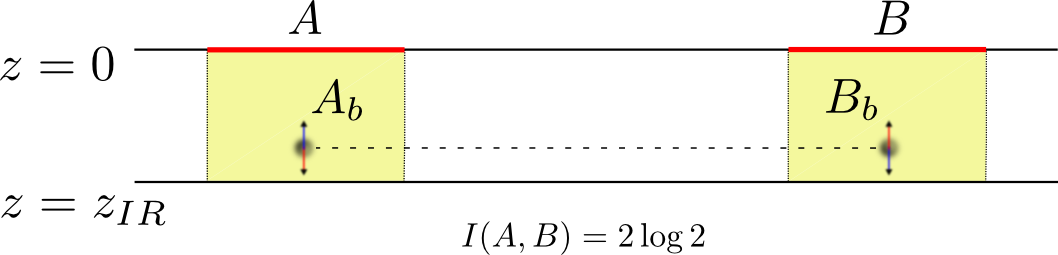}
\vspace{5mm}
\caption{   We consider two regions and their mutual information. In each bulk region we have a
quantum spin. The two spins are in an EPR configuration.    }
\label{EPR}
\end{center}
\end{figure}

Imagine two well separated regions $A$ and $B$ in such a way that their mutual information
vanishes according to the classical RT formula. In the vacuum, the mutual information decays at
long distances.
Here we  add two spins that are EPR correlated as indicated in figure \ref{EPR} . We can imagine these
as arising form the spin of two (fermionic) glueballs in the boundary theory which corresponds
to two particles in the bulk.

In this case the bulk entanglement entropy contains a non-zero
 piece which is independent of the separation,
for large separations. This is just simply the usual mutual information of two spins, $I = 2 \log 2$.
Of course we can consider a more complex system with the same type of result. This
contribution is given by the bulk entanglement term in \nref{finalqu}.



\section*{Acknowledgements}

 We would like to thank  T. Hartman, C. Herzog,  I. Klebanov,  R. Myers and T. Takayanagi
 for discussions.
 JM was supported in part by U.S.~Department of Energy grant DE-SC0009988 . AL acknowledges support from ``Fundacion La Caixa''. TF was supported by  NSF grant PHY 0969448.


\begin{thebibliography}{99}


 \bibitem{RT}
  S.~Ryu and T.~Takayanagi,
  ``Holographic derivation of entanglement entropy from AdS/CFT,''
  Phys.\ Rev.\ Lett.\  {\bf 96}, 181602 (2006)
  [hep-th/0603001].
\bibitem{RT2}
  T.~Nishioka, S.~Ryu and T.~Takayanagi,
  ``Holographic Entanglement Entropy: An Overview,''
  J.\ Phys.\ A {\bf 42}, 504008 (2009)
  [arXiv:0905.0932 [hep-th]].

\bibitem{Faulkner:2013yia}
  T.~Faulkner,
  ``The Entanglement Renyi Entropies of Disjoint Intervals in AdS/CFT,''
  arXiv:1303.7221 [hep-th].

\bibitem{Hartman:2013mia}
  T.~Hartman,
  ``Entanglement Entropy at Large Central Charge,''
  arXiv:1303.6955 [hep-th].
\bibitem{gge}
  A.~Lewkowycz and J.~Maldacena,
  ``Generalized gravitational entropy,''
  arXiv:1304.4926 [hep-th].

\bibitem{Bianchi:2012ev}
  E.~Bianchi and R.~C.~Myers,
  ``On the Architecture of Spacetime Geometry,''
  arXiv:1212.5183 [hep-th].


\bibitem{Srednicki:1993im}
  M.~Srednicki,
  ``Entropy and area,''
  Phys.\ Rev.\ Lett.\  {\bf 71}, 666 (1993)
  [hep-th/9303048].

\bibitem{Bombelli:1986rw}
  L.~Bombelli, R.~K.~Koul, J.~Lee and R.~D.~Sorkin,
  ``A Quantum Source of Entropy for Black Holes,''
  Phys.\ Rev.\ D {\bf 34}, 373 (1986).

\bibitem{Callan:1994py}
  C.~G.~Callan, Jr. and F.~Wilczek,
  ``On geometric entropy,''
  Phys.\ Lett.\ B {\bf 333}, 55 (1994)
  [hep-th/9401072].

\bibitem{Susskind:1994sm}
  L.~Susskind and J.~Uglum,
  ``Black hole entropy in canonical quantum gravity and superstring theory,''
  Phys.\ Rev.\ D {\bf 50}, 2700 (1994)
  [hep-th/9401070].
\bibitem{Solodukhin:1994yz}
  S.~N.~Solodukhin,
  ``The Conical singularity and quantum corrections to entropy of black hole,''
  Phys.\ Rev.\ D {\bf 51}, 609 (1995)
  [hep-th/9407001].
\bibitem{Fursaev:1994ea}
  D.~V.~Fursaev and S.~N.~Solodukhin,
  ``On one loop renormalization of black hole entropy,''
  Phys.\ Lett.\ B {\bf 365}, 51 (1996)
  [hep-th/9412020].

\bibitem{Cooperman:2013iqr}
  J.~H.~Cooperman and M.~A.~Luty,
  ``Renormalization of Entanglement Entropy and the Gravitational Effective Action,''
  arXiv:1302.1878 [hep-th].

\bibitem{Solodukhin.BH}
  S.~N.~Solodukhin,
  ``Entanglement entropy of black holes,''
  Living Rev.\ Rel.\  {\bf 14}, 8 (2011)
  [arXiv:1104.3712 [hep-th]].
\bibitem{offvson}
  V.~P.~Frolov, D.~V.~Fursaev and A.~I.~Zelnikov,
  ``Black hole entropy: Off-shell versus on-shell,''
  Phys.\ Rev.\ D {\bf 54}, 2711 (1996)
  [hep-th/9512184].
\bibitem{Barrella:2013wja}
  T.~Barrella, X.~Dong, S.~A.~Hartnoll and V.~L.~Martin,
  ``Holographic entanglement beyond classical gravity,''
  arXiv:1306.4682 [hep-th].
\bibitem{Iyer:1994ys}
  V.~Iyer and R.~M.~Wald,
  ``Some properties of Noether charge and a proposal for dynamical black hole entropy,''
  Phys.\ Rev.\ D {\bf 50}, 846 (1994)
  [gr-qc/9403028].


 \bibitem{highermyers}
   L.~-Y.~Hung, R.~C.~Myers and M.~Smolkin,
   ``On Holographic Entanglement Entropy and Higher Curvature Gravity,''
   JHEP {\bf 1104}, 025 (2011)
   [arXiv:1101.5813 [hep-th]].
 \bibitem{higherparnachev}
   J.~de Boer, M.~Kulaxizi and A.~Parnachev,
   ``Holographic Entanglement Entropy in Lovelock Gravities,''
   JHEP {\bf 1107}, 109 (2011)
   [arXiv:1101.5781 [hep-th]].
\bibitem{Bhattacharyya:2013jma}
  A.~Bhattacharyya, A.~Kaviraj and A.~Sinha,
  ``Entanglement entropy in higher derivative holography,''
  arXiv:1305.6694 [hep-th].
\bibitem{squashedcones}
  D.~V.~Fursaev, A.~Patrushev and S.~N.~Solodukhin,
  ``Distributional Geometry of Squashed Cones,''
  arXiv:1306.4000 [hep-th].


\bibitem{Sen:2012dw}
  A.~Sen,
  ``Logarithmic Corrections to Schwarzschild and Other Non-extremal Black Hole Entropy in Different Dimensions,''
  arXiv:1205.0971 [hep-th].

\bibitem{Headrick:2007km}
  M.~Headrick and T.~Takayanagi,
  ``A Holographic proof of the strong subadditivity of entanglement entropy,''
  Phys.\ Rev.\ D {\bf 76}, 106013 (2007)
  [arXiv:0704.3719 [hep-th]].



\bibitem{ks}
  I.~R.~Klebanov and M.~J.~Strassler,
  ``Supergravity and a confining gauge theory: Duality cascades and chi SB resolution of naked singularities,''
  JHEP {\bf 0008}, 052 (2000)
  [hep-th/0007191].

\bibitem{Aharony:2000pp}
  O.~Aharony,
  ``A Note on the holographic interpretation of string theory backgrounds with varying flux,''
  JHEP {\bf 0103}, 012 (2001)
  [hep-th/0101013].
\bibitem{Gubser:2004qj}
  S.~S.~Gubser, C.~P.~Herzog and I.~R.~Klebanov,
  ``Symmetry breaking and axionic strings in the warped deformed conifold,''
  JHEP {\bf 0409}, 036 (2004)
  [hep-th/0405282].

\bibitem{ent.and.confinement}
  I.~R.~Klebanov, D.~Kutasov and A.~Murugan,
  ``Entanglement as a probe of confinement,''
  Nucl.\ Phys.\ B {\bf 796}, 274 (2008)
  [arXiv:0709.2140 [hep-th]].

\bibitem{Fujita:2009kw}
  M.~Fujita, W.~Li, S.~Ryu and T.~Takayanagi,
  ``Fractional Quantum Hall Effect via Holography: Chern-Simons, Edge States, and Hierarchy,''
  JHEP {\bf 0906}, 066 (2009)
  [arXiv:0901.0924 [hep-th]].

\bibitem{Solodukhin:2008dh}
  S.~N.~Solodukhin,
  ``Entanglement entropy, conformal invariance and extrinsic geometry,''
  Phys.\ Lett.\ B {\bf 665}, 305 (2008)
  [arXiv:0802.3117 [hep-th]].
\bibitem{Headrick:2010zt}
  M.~Headrick,
  ``Entanglement Renyi entropies in holographic theories,''
  Phys.\ Rev.\ D {\bf 82}, 126010 (2010)
  [arXiv:1006.0047 [hep-th]].

\bibitem{raams1}
  M.~Van Raamsdonk,
  ``Comments on quantum gravity and entanglement,''
  arXiv:0907.2939 [hep-th].
\bibitem{raams2}
  M.~Van Raamsdonk,
  ``Building up spacetime with quantum entanglement,''
  Gen.\ Rel.\ Grav.\  {\bf 42}, 2323 (2010)
  [Int.\ J.\ Mod.\ Phys.\ D {\bf 19}, 2429 (2010)]
  [arXiv:1005.3035 [hep-th]].

\bibitem{ciracbound}
M.M. Wolf, F. Verstraete, M.B. Hastings, J.I. Cirac,
``Area laws in quantum systems: mutual information and correlations,''
Phys. Rev. Lett. 100, 070502 (2008), [arXiv:0704.3906[quant-ph]]


\bibitem{Casini:2008wt}
  H.~Casini and M.~Huerta,
  ``Remarks on the entanglement entropy for disconnected regions,''
  JHEP {\bf 0903}, 048 (2009)
  [arXiv:0812.1773 [hep-th]].


\bibitem{Calabrese:2010he}
  P.~Calabrese, J.~Cardy and E.~Tonni,
  ``Entanglement entropy of two disjoint intervals in conformal field theory II,''
  J.\ Stat.\ Mech.\  {\bf 1101}, P01021 (2011)
  [arXiv:1011.5482 [hep-th]].
\bibitem{Cardy:2013nua}
  J.~Cardy,
  ``Some Results on Mutual Information of Disjoint Regions in Higher Dimensions,''
  arXiv:1304.7985 [hep-th].






\bibitem{wilsonloopMI}
  J.~Molina-Vilaplana,
  ``On the Mutual Information between disconnected regions in AdS/CFT,''
  arXiv:1305.1064 [hep-th].



\bibitem{Gubser:1998bc}
  S.~S.~Gubser, I.~R.~Klebanov and A.~M.~Polyakov,
  ``Gauge theory correlators from noncritical string theory,''
  Phys.\ Lett.\ B {\bf 428}, 105 (1998)
  [hep-th/9802109].

\bibitem{Witten:1998qj}
  E.~Witten,
  ``Anti-de Sitter space and holography,''
  Adv.\ Theor.\ Math.\ Phys.\  {\bf 2}, 253 (1998)
  [hep-th/9802150].




\bibitem{arakilieb}
H.~Araki and E.~Lieb,
``Entropy inequalities,'' Commun.~Math.~Phys.~{\bf 18} (1970) 160�170.


\bibitem{Hubeny:2013gta}
  V.~E.~Hubeny, H.~Maxfield, M.~Rangamani and E.~Tonni,
  ``Holographic entanglement plateaux,''
  arXiv:1306.4004 [hep-th].


\bibitem{Blanco:2013joa}
  D.~D.~Blanco, H.~Casini, L.~-Y.~Hung and R.~C.~Myers,
  ``Relative Entropy and Holography,''
  arXiv:1305.3182 [hep-th].

\bibitem{Wolf:2006zzb}
  M.~M.~Wolf,
  ``Violation of the entropic area law for Fermions,''
  Phys.\ Rev.\ Lett.\  {\bf 96}, 010404 (2006)
  [quant-ph/0503219].

 \bibitem{Gioev}
D. Gioev and I.  Klich 
 ``Entanglement entropy of fermions in any dimension and the Widom conjecture'',
 	Phys. Rev. Lett. 96, 100503 (2006), [quant-ph/0504151].
 

\bibitem{Faulkner:2009wj}
  T.~Faulkner, H.~Liu, J.~McGreevy and D.~Vegh,
  ``Emergent quantum criticality, Fermi surfaces, and AdS(2),''
  Phys.\ Rev.\ D {\bf 83}, 125002 (2011)
  [arXiv:0907.2694 [hep-th]].


\bibitem{Hartnoll:2010gu}
  S.~A.~Hartnoll and A.~Tavanfar,
  ``Electron stars for holographic metallic criticality,''
  Phys.\ Rev.\ D {\bf 83}, 046003 (2011)
  [arXiv:1008.2828 [hep-th]].


\bibitem{Sachdev:2011ze}
  S.~Sachdev,
  ``A model of a Fermi liquid using gauge-gravity duality,''
  Phys.\ Rev.\ D {\bf 84}, 066009 (2011)
  [arXiv:1107.5321 [hep-th]].

\bibitem{Ogawa:2011bz}
  N.~Ogawa, T.~Takayanagi and T.~Ugajin,
  ``Holographic Fermi Surfaces and Entanglement Entropy,''
  JHEP {\bf 1201}, 125 (2012)
  [arXiv:1111.1023 [hep-th]].




  \end{thebibliography}
\end{document}